# Optimizing Surveillance Satellites for the Synthetic Theater Operations Research Model

Steven M. Warner, Major, USMC
The Joint Staff, J8 Warfighting Analysis Division, The Pentagon, 2B852A
steven.m.warner.mil@mail.mil

Johannes O. Royset
Operations Research Department
Naval Postgraduate School, Monterey CA
joroyset@nps.edu

The Synthetic Theater Operations Research Model (STORM) simulates theater-level conflict and requires inputs about utilization of surveillance satellites to search large geographical areas. We develop a mixed-integer linear optimization model that prescribes plans for how satellites and their sensors should be directed to best search an area of operations. It also specifies the resolution levels employed by the sensors to ensure a suitable fidelity of the resulting images. We solve large-scale instances of the model involving up to 22 million variables and 11 million constraints in scenarios derived from STORM. On average, the model yields 55% improvement in search coverage relative to an existing heuristic algorithm in STORM.

Areas and Methods:  Search and detection; Combat modeling; Intelligence, surveillance, and reconnaissance; Integer programming

## INTRODUCTION

Military organizations strive to search large geographic areas for enemy combatants and their support units while making best use of assets such as satellites. This task is challenging and easily results in large, unsearched regions and inefficient use of assets. In this paper, we address the *look allocation problem* where an area of operations is divided into *grid cells*, with each grid cell having a *priority penalty* indicative of how often it needs to be searched. The planner has sensors on satellites that orbit the earth. During each pass over the area, a sensor can look for objects along a two-dimensional strip on the ground, which we refer to as a *swath*. However, only a portion of a swath can be looked at if the sensor were to produce high-resolution images. A planner would need to make a constrained decision about which grid cells within a swath should be looked at and with what resolution, while minimizing the priority penalties that tend to increase over time for grid cells ignored and reset to zero for those looked at. Moreover, each grid cell should receive at least one look during the planning horizon. The size of the area of operations, the multitude of sensors and resolution levels, as well as the large number of satellite passes over the area complicate the problem.

We study the problem within the context of the campaign simulation tool Synthetic Theater Operations Research Model (STORM), which furnishes realistic problem instances as well as a benchmark for comparison (Group W, 2005). An existing heuristic algorithm for allocating looks within STORM tends to over-allocate to certain grid cells and under-allocate to others, use too high or too low resolution levels, or adopt too long or too short revisit intervals. The heuristic can leave large areas unsearched. This motivates our development of the *Look Optimization Model* (LOM), which is a mixed-integer linear program for optimally allocating looks during a planning horizon.

The look allocation problem shares similarities with the orienteering problem (OP) (Tsiligirides, 1984), where competitors use land navigation to race to visit checkpoints. Competitors seek to maximize their score by visiting as many checkpoints as possible within the allotted time, often referred to as a score orienteering event. In its most basic form, OP combines node selection with shortest paths and relates to both the knapsack and travelling salesman problems (Gunawan et al., 2016). OP applies to challenges in logistics, vehicle routing, tourism, and military operations research (Vansteenwegen et al., 2011).

There are numerous variants of OP: Team OP seeks to find multiple paths that maximize a collective score, time-constrained OP incorporates time windows for specific node visitation, and generalized OP incorporates a nonlinear objective function for computing the overall score of a competitor (Wang et al., 1995; Geem et al., 2005; Wang et al., 2008; Royset and Reber, 2009; Vansteenwegen et al., 2011). Ilhan et al. (2008) study OP with stochastic profits where nodes are assigned normally distributed random profits with the goal of exceeding a certain profit threshold. Gendreau et al. (1998) incorporate compulsory vertices into OP which forces requirements on certain checkpoints to be visited. Fomin and Lingas (2002) consider the time-dependent OP where the travel time between checkpoints depends upon the competitors start time. The capacitated OP limits the total number of checkpoints that each competitor can visit (Archetti et al., 2009). Pietz and Royset (2013) introduce varying quantities of rewards at each node which depend upon resource expenditure to reach a specific node. The multi-profit OP varies profit by time spent at each checkpoint and requires selecting both checkpoint sequence and time spent at each checkpoint (Kim et al., 2020). Each of these variants is NP-hard. Computational approaches include efforts to reduce runtimes through specialized branch-and-bound algorithms (Pietz and Royset, 2013), neural networks (Wang et al., 1995, 2008), and heuristics (Kim et al., 2020; Archetti et al., 2009; Gendreau et al., 1998).

The maximum covering location problem (MCLP) is a thoroughly studied NP-hard problem. Given sets with a nonempty intersection, we seek to choose a limited number of the sets that maximize the total coverage of elements (Church and ReVelle, 1974). Computational solution approaches include branch-and-cut algorithms (Cordeau et al., 2019; Vasta Jayaswal, 2021), greedy heuristics (Sekar Amarilies et al., 2020; Seyhan et al., 2018), metaheuristics (Pereira et al., 2015), minmax regret models (Baldomero-Naranjo et al., 2021), and simulated annealing (Zarandi et al., 2013); see also Hoogendoorn (2017) for a review. In the context of counterdrug trafficking, Price et al. (2022) extend MCLP to account for multiple drug cartels and counterdrug assets and formulate three mixed-integer programs. However, their programs only model a small subset of operationally relevant constraints.

Optimal search theory (Stone et al., 2016) centers on allocating search effort (resources) to maximize the probability of detecting a target; see Ding (2018) for a recent review. Typically, the target is moving according to a probability distribution resulting in the need for specialized branch-and-bound algorithms (Dell et al., 1996; Washburn, 1998; Lau et al., 2008; Bourque, 2019; Sato and Royset, 2010), cutting plane methods (Royset and Sato, 2010), open-loop network flow algorithms (Berger et al., 2021), and heuristics (Dell et al., 1996; Abi-Zeid et al., 2019).

The look allocation problem of the present paper deviates from the above problems due to the specific requirements of STORM. There is no specific target to detect as in classical search problems, but rather a desire to look sufficiently often in each grid cell, with importance given according to priority penalties. There is also a need to explicitly model the trade-off between an accurate look in a small number of grid cells and a coarse look covering many grid cells. The satellite orbits, assumed given, specify timing and location of swaths and thus constrain where and when looks are available. However, there is no "travel time" between grid cells, which is central to all variants of OP. The study by Peng et al. (2020) appears to be the most related to our setting. It examines how to manipulate the look angle by a satellite to detect targets with varying rewards and time windows. However, it only considers a single orbit and does not account for the time since the last look in a grid cell, which is central to the assessment of surveillance plans in STORM. It also relies on the solution of a dynamic program, which is challenging at the scale needed for STORM campaign analysis; 1000 grid cells and 50 swaths are common. Thus, we find it difficult to directly expand on the earlier efforts and instead develop a mixed-integer linear optimization model tailored to the needs of STORM. The paper derives from the Master's thesis by the first author (Warner, 2022).

We proceed in the next section with a review of STORM and how it defines the look allocation problem. Section 3 presents the mixed-integer linear optimization model. Section 4 discusses several computational studies. The paper ends with conclusions in Section 5.

## STORM AND SCENARIO BACKGROUND

STORM simulates theater-level conflict while representing all warfighting domains across the range of military operations. In this study, we use the unclassified, modern STORM scenario titled Punic21 which has two combatants, Blue and Red. Thus, the scenario gives rise to two instances of the look allocation problem: planning Blue's looks at Red and planning Red's looks at Blue. Punic21 takes place in the geographic regions of Europe and North Africa; see Group W (2019a,d,c,b). The area of operations is divided into 14,400 grid cells with size 50 km by 50 km covering 36 million square km. Typically, about 10% of these grid cells are of direct interest to a combatant in a particular case. Other scenarios have similar characteristics.

Punic21 has four Reconnaissance and one Space Based LASER (SBL) satellites following a dual body Kepler orbit for each combatant. STORM produces satellite entry and exit coordinates over the area of operations and corresponding times. The satellites make 4 to 6 passes over the area of interest every 24-hour period. Each pass of each satellite produces a swath. The Reconnaissance satellites have four sensors. SBL has three sensors. Sensors are named after the type of functions they perform which are electro-optical imagery analysis (IMINT), infrared day IMINT, infrared night IMINT, and synthetic aperture radar IMINT.

Each sensor has a limited number of looks it can allocate to grid cells per swath which is determined by the adopted resolution levels. These resolutions follow the National Imagery Interpretability Rating Scale (NIIRS) and have at most nine discrete levels. A higher resolution (as measured by NIIRS) equates to a more detailed look and an increased likelihood of detecting an object in a grid cell. However, setting a high resolution limits the number of looks to distribute per swath. Each sensor and resolution level is associated with a square area budget and a look count budget. Tables 1 and 2 give these budgets as functions of resolution for electro-optical sensors and other sensors, respectively. Since each grid cell is 2,500 square km (in this scenario), the second column of the tables also computes the number of grid cells corresponding to a specific square area budget.

| Resolution | Square area budget | Look count budget | Cost factor |
|---|---|---|---|
| 1 | 500,000 sq km = 200 grid cells | 100 | 0.01 |
| 4 | 10,000 sq km = 4 grid cells | 90 | 0.25 |
| 9 | 500 sq km = 0.2 grid cells | 20 | - |

**Table 1. Data for electro-optical sensors**. Square area budget, look count budget, and cost factors for electro-optical sensors as functions of resolution (Group W, 2005).

| Resolution | Square area budget | Look count budget | Cost factor |
|---|---|---|---|
| 1 | 500,000 sq km = 200 grid cells | 100 | 0.01 |
| 2 | 100,000 sq km = 40 grid cells | 100 | 0.025 |
| 3 | 50,000 sq km = 20 grid cells | 90 | 0.05 |
| 4 | 10,000 sq km = 4 grid cells | 90 | 0.25 |
| 5 | 5,000 sq km = 2 grid cells | 80 | 0.5 |
| 6 | 1,000 sq km = 0.4 grid cells | 80 | - |
| 7 | 500 sq km = 0.2 grid cells | 70 | - |
| 8 | 100 sq km = 0.04 grid cells | 50 | - |
| 9 | 50 sq km = 0.02 grid cells | 40 | - |

**Table 2. Data for other sensors**. Square area budget, look count budget, and cost factors for infrared day, infrared night, and synthetic aperture radar sensors as functions of resolution (Group W, 2005).

Since neither budget should be exceeded, it is the smaller number in columns two and three of each row of Tables 1 and 2 that becomes limiting. For example, an electro-optical sensor at resolution 1 can look in 100 grid cells. At resolution 4, that number is reduced to 4; see Table 1. To account for the fact that a sensor can vary the resolution during a swath, we convert these numbers into cost factors representing the fraction of the budget used to search a single grid cell with a particular resolution. For example, the cost to look at one grid cell by an electro-optical sensor at resolution 1 is then $1/100 = 0.01$, while at resolution 4 the number is $1/4 = 0.25$. The last columns of Tables 1 and 2 give the resulting costs. This means a swath by an electro-optical sensor can involve looking at 50 grid cells at resolution 1, which costs $50 \cdot 0.01 = 0.50$, and looking at two grid cells with resolution 4, which costs $2 \cdot 0.25 = 0.50$, and still satisfying the budget of 1. Of course, there are many other feasible combinations. Resolution levels producing cost factors greater than 1 are omitted as they automatically will exceed the budget of 1. This omission is reasonable as large area search would not involve such high resolutions.

STORM assigns importance to each grid cell through a priority penalty, which in turn is defined by 13 parameters. Figure 1 depicts the priority penalties for three grid cells and how they evolve over time. The white curve corresponds to grid cell 1. It starts at zero, but as time advances since the last look in grid cell 1 the penalty increases. For example, if the next swath occurs after 20 hours but grid cell 1 is not looked at, then a penalty of 0.01 is incurred. If a second swath occurs at 37 hours and again the grid cell is not looked at, then an additional penalty of 0.05 is incurred. These penalties are relatively small, but if grid cell 1 is consistently ignored, then the penalty increases dramatically as seen from the steep growth of the white curve. Grid cell 2 is assigned priority penalty according to the dark curve in Figure 1. This grid cell is more "important" because its priority penalty increases much faster than for grid cell 1. In fact, if grid cell 2 receives no look at 20 hours, then a penalty of 0.42 would have occurred. However, grid cell 2 is indeed looked at and this resets the priority penalty to zero. Figure 1 also depicts the priority penalties of grid cell 3 in light gray. While grid cell 3 is not looked at during the first swath (20 hours), which causes a penalty of 0.08, it receives a look at 37 hours. This resets the priority penalty for grid cell 3 to zero.

STORM has a heuristic algorithm for solving the look allocation problem. For each swath, it allocates looks by examining all cells covered by the swath, assigning looks in order of decreasing priority penalty until the budget of looks is exhausted. In the case of a tie between two grid cells, the heuristic assigns the look to the grid cell it encounters first when proceeding in a north-to-south, west-to-east manner. The resolution level needs to be determined by the user. This greedy algorithm does not solve the look allocation problem and in particular fails to ensure that all grid cells receive at least one look during the planning horizon. Its performance is also highly dependent on a "good" choice of resolution, which is difficult to determine.

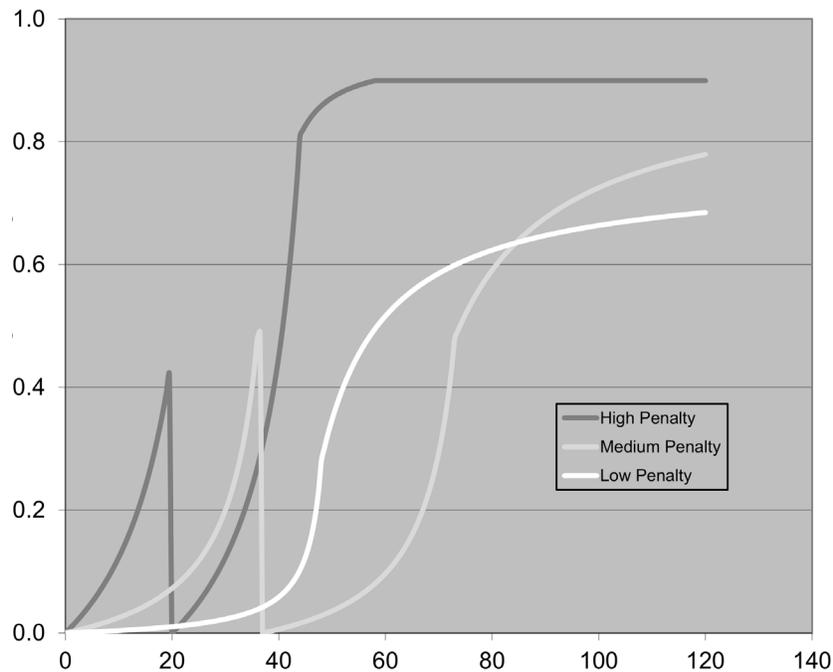

**Figure 1.** Each line depicts priority penalties for a grid cell. Time is on the first axis (hours) and the associated priority penalty is on the second axis.

## LOM FORMULATION

In this section, we formulate the mixed-integer optimization model LOM for allocating looks to grid cells.

A feature of LOM is its handling of "time." We do not discretize time into time-steps but rather adopt uneven jumps in time corresponding to the occurrence of each swath. Since satellite passes over the area of operations take place at known points in time, swath number *s* corresponds to a specific point in time. Thus, a swath counter can serve as our "unit" of time. A satellite may have several sensors, which then produce a corresponding number of swaths. These swaths occur at the same time, and we order them arbitrarily. We use the term *gap* to denote the number of swaths since the last look in a grid cell.

In addition to minimizing priority penalties accumulated during a planning horizon, we also (strongly) encourage at least one look in each grid cell and allow for a user-specified threshold resolution. LOM takes the following form.

**Model Index Use** (denoted by subscripts) [~cardinality]

| | |
|---|---|
| $c = 1, \dots, C$ | Grid cell [~1,000] |
| $r = 1, \dots, R$ | Resolution [~5] |
| $s = 0, 1, \dots, S$ | Swath [~40] |
| $g = 0, 1, \dots, G$ | Gap [~40] |

**Given data** [units]

| | |
|---|---|
| $pen_{c,s,g}$ | Penalty in grid cell $c$ for swath $s$ when the gap is $g$ [unitless] |
| $cost_{s,r}$ | Cost of a look at resolution $r$ for swath $s$ [unitless] |
| $rmin_c$ | Threshold resolution for looks in grid cell $c$ [unitless] |
| $maxlow$ | Maximum number of looks with resolution less than $rmin_c$ [unitless] |
| $never$ | Penalty for never looking at a grid cell [unitless] |
| $bigM$ | Large constant for significant swath gaps between looks [unitless] |

**Decision Variables** [units]

| | |
|---|---|
| $X_{c,s,r}$ | 1 if grid cell $c$ receives a look with resolution $r$ from swath $s$; 0 otherwise [binary] |
| $Y_{c,s,g}$ | 1 if at swath $s$ there is a gap $g$ since last look in grid cell $c$; 0 otherwise [binary] |
| $G_{c,s}$ | At swath $s$, gap since last look in grid cell $c$ |
| $Z_c$ | Amount by which the at-least-once constraint is violated in grid cell $c$ |

**Formulation**

$$\text{minimize} \sum_{c,s,g} pen_{c,s,g} \, Y_{c,s,g} + never \sum_{c} Z_c \qquad (1)$$

subject to

$$\sum_{g} g \, Y_{c,s,g} = G_{c,s} \qquad\qquad \forall c,s \qquad (2)$$

$$\sum_{g} Y_{c,s,g} = 1 \qquad\qquad \forall c,s \qquad (3)$$

$$1 + G_{c,s-1} \leq G_{c,s} + bigM \sum_{r} X_{c,s,r} \qquad \forall c,s > 0 \qquad (4)$$

$$\sum_{s,r} X_{c,s,r} \geq 1 - Z_c \qquad\qquad \forall c \qquad (5)$$

$$\sum_{c,r} cost_{s,r} \, X_{c,s,r} \leq 1 \qquad\qquad \forall s \qquad (6)$$

$$\sum_{s,r \, < \, rmin_c} X_{c,s,r} \leq maxlow \qquad\qquad \forall c \qquad (7)$$

$$X_{c,0,r} = 0 \qquad\qquad \forall c,r \qquad (8)$$

$$Y_{c,s,g} = 0 \qquad\qquad \forall c,s < g \qquad (9)$$

$$0 \leq G_{c,s} \qquad\qquad \forall c,s \qquad (10)$$

$$G_{c,0} = 0 \qquad\qquad \forall c \qquad (11)$$

$$0 \leq Z_c \leq 1 \qquad\qquad \forall c \qquad (12)$$

**Discussion**

The first term in the objective function (1) sums up the priority penalties incurred during the planning horizon. The gap index $g$ counts the number of swaths that has passed since a cell was looked at. Thus, $pen_{c,s,g}$ can be precomputed by leveraging the known time of each swath; see Figure 1 for typical values.

The second term assigns an additional penalty $never$ for each grid cell never looked at. Since the look allocation problem seeks a plan with looks in all grid cells, $never$ is usually set to a high number. However, we shy away from insisting on $Z_c = 0$ in (5) because practical instances of the model tend to include "difficult" grid cells not accessible by any swath or simply too many grid cells relative to the number of looks available. It is therefore more meaningful to utilize an elastic implementation of the "at-least-one-look" requirement.

We use constraint (2) to maintain the value of the gap between looks in a grid cell. Constraint (3) ensures that there is a unique gap since the last look in a grid cell. Constraint (4) allows the gap to reset to zero when a look occurs and forces an increase in the gap otherwise. We choose $bigM$ as small as possible, i.e., $bigM = S$, the number of swaths. Constraint (5) specifies the desired number of looks per grid cell during the planning horizon; here set to 1 but this can trivially be changed. Failure to comply with this constraint produces a penalty in the objective via the variable $Z_c$. The budget constraint (6) ensures that only available looks are allocated as permitted by their costs. Constraint (7) limits the number of looks with low resolution to $maxlow$. If $maxlow = 1$, then one look is permitted to happen below the threshold resolution $rmin_c$; all other looks shall use the threshold resolution or higher. The constraint prevents any such low resolution looks when $maxlow = 0$. The remaining constraints (8) through (12) set certain variables to zero and specify upper and lower bounds for variables.

**RESULTS FROM CASE STUDIES**

We are interested in both the computational and operational results of LOM. Computational results address concerns about run times and identify effective solver options. Operational results focus on grid cell coverage and comparison with the STORM heuristic while adjusting the number of swaths and threshold resolution.

Our base case involves two types of grid cells (high priority and low priority) with priority penalties according to pre-loaded values in STORM; details omitted. If not otherwise stated, we set the number of grid cells to $C = 1,415$ (1,158 low priority and 257 high priority), the number of resolution levels $R = 5$, the number of swaths $S = 34 = G$, $maxlow = 0$, $never = 100,000$, and $rmin_c = 4$ for all grid cells. The 34 swaths correspond to a planning horizon of 12 hours. We use the solver CPLEX version 20.1.0.0 (CPLEX, IBM Ilog, 2017). The optimality tolerance is 5%.

**Computational Results**

Using a laptop with 16 GB RAM and a 2.69 GHz processor running Windows 10 Pro and a desktop computer with 128 GB of RAM and a 2.10 GHz processor running Windows 10 Enterprise, we examine the effect of increasing variables and constraints by varying grid cells and swaths; see Table 3. Fixing the number of grid cells at 100, we recognize from rows 1-8 that the runtime grows significantly with the number of swaths $S$. While the optimality tolerance is 5%, the solver tends to produce significantly better solutions as reported in column 6. The availability of additional RAM appears beneficial in reducing runtimes as the 128-GB desktop consistently outperforms the 16-GB laptop. Rows 9-13 of Table 3 report results for instances with more grid cells. We note that the performance of the solver occasionally improves with larger instances; see rows 10 and 13 where an instance with 1,415 grid cells solves but a similar instance with 500 gird cells fails because the solver runs out of memory.

| $C$ | $S$ | Number of variables | Number of constraints | Runtime (minutes) | Optimality gap | RAM (GB) |
|---|---|---|---|---|---|---|
| 100 | 16 | 39,201 | 24,638 | 0.26 | 0% | 16 |
| 100 | 18 | 47,601 | 29,885 | 0.57 | 0% | 16 |
| 100 | 42 | 210,801 | 117,899 | 57.46 | 0.04% | 16 |
| 100 | 42 | 210,801 | 117,899 | 38.96 | 0.51% | 128 |
| 100 | 46 | 249,201 | 139,303 | 218.11 | 0.96% | 16 |
| 100 | 46 | 249,201 | 139,303 | 63.53 | 2.00% | 128 |
| 100 | 65 | 475,301 | 259,547 | 190.32 | 1.79% | 16 |
| 100 | 65 | 475,301 | 259,547 | 184.61 | 3.70% | 128 |
| 200 | 50 | 581,601 | 323,797 | 43.94 | 0.01% | 16 |
| 500 | 52 | 1,508,501 | 839,178 | 532.69 | failed | 16 |
| 1,415 | 18 | 673,541 | 421,160 | 4.39 | 0% | 16 |
| 1,415 | 21 | 873,056 | 539,778 | 6.38 | 0% | 16 |
| 1,415 | 51 | 4,269,056 | 2,373,438 | 49.09 | 0% | 16 |

**Table 3. Runtimes of grid cell and swath variations.** Minutes of runtime to satisfy 5% optimality tolerance and the actual optimality gap achieved for CPLEX by varying number of cells $C$ and swaths $S$. The entry "failed" indicates that no feasible solution is achieved before memory exhausted.

The lengthy runtimes in Table 3 are caused by difficulties with finding a feasible solution. For instance, row 3 and row 10 of the table found initial integer solutions at node 18,983 and 3,565 in the branch-and-bound tree, respectively. After producing an integer solution, CPLEX quickly works towards a near-optimal solution.

To reduce runtime, we experiment with three solver options: node selection strategy, rounding of priority penalty coefficients, and warm start.

The first option, best estimate search (BES), selects the next node in the branch-and-bound tree through a ratio of the previous node's progress toward a feasible integer solution compared with the objective function value. Changing the option to BES can be helpful in producing feasible solutions.

The second option implements rounding of the priority penalties (RPP) to the sixth decimal place. This implies a practically insignificant change to the priority penalties, with expected savings on memory and computing time.

The third option supplies CPLEX with the worst possible feasible solution of no looks occurring over the entire planning horizon to implement as a warm start (WS). Specifically, we set $Z_c = 1$, $G_{c,s} = s, X_{c,s,r} = 0$ for all $c, s,$ and $r$. Additionally, we assign $Y_{c,s,g} = 1$ for all $s = g$ and 0 otherwise. This solution immediately achieves an optimality gap of 100% and may subsequently permit quicker development in cutoff values for the best integer solution as well as enhance probing to fix variables (Klotz and Newman 2013). Table 4 summarizes our findings when using $rmin_c = 4$ and $rmin_c = 1$ for high- and low-priority grid cells, respectively.

| $S$ | Number of variables | Number of constraints | No option | BES option | RPP option | WS option | BES, RPP & WS options |
|---|---|---|---|---|---|---|---|
| 10 | 266,021 | 184,824 | 1.88 [1.11] | 1.85 [0.03] | 0.98 [0.016] | 0.63 [0.01] | 0.71 [3.54] |
| 22 | 945,221 | 585,761 | 96.15 [1.60] | 123.79 [0.18] | 101.27 [1.69] | 1.87 [0.03] | 1.94 [0.6] |
| 34 | 2,031,941 | 1,183,678 | 1,466.37 [failed] | 806.96 [2.91] | 676.63 [0.55] | 4.05 [1.52] | 4.15 [1.04] |

**Table 4. Runtimes (minutes) for different CPLEX options.** Minutes of runtime on 128-GB desktop to satisfy 5% optimality tolerance and, in brackets, the actual optimality gap (%) achieved.

Table 4 highlights that WS provides a significant reduction in runtime whereas BES and RPP sometimes improve upon the default setting. Column eight shows that all three options combined do not significantly detract from WS. Table 5 explores the choice of options further on larger instances. Again, we conclude that WS is highly beneficial and may also be combined with other options.

| $S$ | Number of variables | Number of constraints | WS option | BES & WS options | RPP & WS options | BES, RPP & WS options |
|---|---|---|---|---|---|---|
| 86 | 11,450,181 | 6,150,850 | 23.13 [2.90] | 23.31 [2.90] | 24.50 [3.70] | 24.68 [3.70] |
| 111 | 18,702,056 | 9,890,060 | 40.51 [4.08] | 40.87 [4.08] | 42.22 [0.70] | 42.14 [4.08] |
| 121 | 22,098,056 | 11,643,435 | 39.55 [9.55] | 39.05 [9.55] | 44.29 [11.06] | 49.60 [11.06] |

**Table 5.  Runtimes (minutes) for different CPLEX options; large instances.**  Minutes of runtime on 128-GB desktop to satisfy 5% optimality tolerance and, in brackets, the actual optimality gap (%) achieved. All experiments in row three exhaust the memory before reaching the 5% optimality tolerance.

Table 6 shows that constraint (7) significantly affects the runtime. In row one, $rmin_c = 1$ for all $c$, which voids constraint (7); this increases the runtimes dramatically compared to the other rows involving $rmin_c > 1$. In fact, the runtimes in row one average 966 minutes compared to 4.2 minutes in the other rows.

| $rmin_c$ | WS | BES & WS | RPP & WS | BES, RPP & WS |
|---|---|---|---|---|
| 1 | 1,186.40 [0.92] | 1,209.12 [0.7] | 690.79 [1.08] | 778.67 [1.1] |
| 2 | 4.10 [0.01] | 4.08 [0.01] | 4.42 [0.01] | 4.57 [0.01] |
| 3 | 4.29 [0.88] | 4.22 [0.88] | 4.27 [0.88] | 4.47 [0.88] |
| 4 | 3.99 [0.08] | 3.81 [0.08] | 4.09 [0.08] | 4.13 [0.08] |
| 5 | 4.14 [0.08] | 4.20 [0.08] | 4.16 [0.08] | 4.15 [0.08] |

**Table 6.  Runtimes (minutes) for varying threshold resolution and CPLEX options.** Minutes of runtime on 128-GB desktop to satisfy 5% optimality tolerance and, in brackets, the actual optimality gap (%) achieved. We require all grid cells to use the specified $rmin_c$ except in the last row ($rmin_c = 5$). In that case, we set $rmin_c = 5$ for the high-priority grid cells and $rmin_c = 4$ for the low-priority grid cells to allow for use of sensors that do not have resolution 5.

If we permit a single look for each grid cell to use *any* resolution, i.e., setting $maxlow = 1$, we also see an increase in computing times; consult Table 7. We observe that BES and WS tend to slightly increase runtime; however, bundling all three options produces relatively low runtimes.

We conclude that it is best to invoke all three options when solving LOM. The remaining results utilize all three options.

| $S$ | WS | BES & WS | RPP & WS | BES, RPP, & WS |
|---|---|---|---|---|
| 22 | 2.00   [0.91] | 1.98   [0.91] | 2.08   [0.91] | 2.11   [0.91] |
| 86 | 1,562.26 [50.21] | 1,581.48 [50.21] | 1,562.92 [49.13] | 1,506.02 [49.13] |

**Table 7. Runtimes (minutes) for $maxlow = 1$.** Minutes of runtime on 128-GB desktop to satisfy 5% optimality tolerance and, in brackets, the actual optimality gap (%) achieved. All experiments have $rmin_c = 2$.

## Operational Results

We next turn to a comparison between allocation plans from LOM and those from the STORM heuristic. Planners aspire to have all (relevant) grid cells looked at once. LOM accounts for this concern via (5), while the heuristic has no direct means of ensuring such a constraint. In the following, we report *coverage* as the fraction of grid cells receiving one look or more relative to the total number of grid cells of interest for a combatant. We focus on adjusting two parameters, the number of swaths, which corresponds to the planning horizon, and the threshold resolution $rmin_c$.

Table 8 shows coverage percentage of the plans produced by LOM and by the STORM heuristic. Naturally, in the presence of few swaths, both methods produce low coverage because there are few looks to allocate. LOM gains an advantage when there are more swaths and thus more looks. In the last row, LOM produces a plan that looks at twice as many grid cells compared to the heuristic. We notice that the number of looks allocated by LOM usually is different than that by the STORM heuristic. This is caused by the fact that LOM optimally selects resolution levels and consequently the number of looks available, while the heuristic utilizes a user-specified resolution.

Table 8 also provides details about allocation between low-priority and high-priority grid cells. Columns 2 and 3 give the numbers of unique grid cells of the high-priority and low-priority kinds, respectively, looked at when using the LOM plans. Parallel results for the STORM heuristic are given by Columns 5 and 6. The rows with 22, 34, and 111 swaths report cases where LOM assigns looks to low-priority grid cells but not to high-priority ones. This is caused by the large value of the parameter *never*, which effectively drives the model to look at as many grid cells as possible. In these instances, low-priority grid cells are simply more accessible and high-priority grid cells are not easily looked at without sacrificing the overall number of grid cells covered. (Since solutions are obtained with 5% optimality tolerance, additional grid cells of either kind might be added for an optimal solution.) Trivially, one can make the parameter *never* in LOM depend on the grid cell. This would afford planners with the flexibility of penalizing a missing high-priority grid cell more severely than a low-priority one.

The last row of Table 8 shows that LOM produces a plan that looks at 409 (unique) grid cells using 416 looks. In contrast, the STORM heuristic looks at 193 grid cells using 455 looks. The STORM heuristic wastes many looks by revisiting grid cells too frequently.

| | LOM | | | STORM heuristic | | |
|---|---|---|---|---|---|---|
| $S$ | High priority | Low priority | Coverage | High priority | Low priority | Coverage |
| 10 | 24 | 16 | 2.83% | 26 | 14 | 2.83% |
| 22 | 0 | 83 [87] | 5.87% | 46 [64] | 28 | 5.23% |
| 34 | 0 | 135 | 9.54% | 55 [76] | 64 | 8.41% |
| 43 | 65 | 106 [107] | 12.08% | 60 [85] | 86 [87] | 10.32% |
| 86 | 100 | 231 [244] | 23.39% | 70 [164] | 105 [171] | 12.37% |
| 111 | 0 | 409 [416] | 28.90% | 77 [228] | 116 [227] | 13.64% |

**Table 8. LOM vs STORM coverage**. LOM compared to STORM coverage (%) and how the coverage distributes between low- and high-priority grid cells (in number of cells looked at of each kind). Numbers in brackets give total number of looks if higher than the number of grid cells looked at.

When we relax $rmin_c$ from 4 to 1 for low-priority grid cells, coverage improves and LOM reaches numbers above 95%; see Table 9. The advantage of LOM over the STORM heuristic is especially significant for moderate planning horizons (i.e., 22-43 swaths). As in Table 8, the solver finds it difficult to identify plans with many looks in high-priority grid cells. If of concern to a planner, this can be addressed by tightening the optimality tolerance and/or adjusting the parameter *never* as discussed above.

| $S$ | LOM | | | STORM heuristic | | |
|---|---|---|---|---|---|---|
| | High priority | Low priority | Coverage | High priority | Low priority | Coverage |
| 10 | 2 | 802 [816] | 56.82% | 26 | 260 [300] | 20.21% |
| 22 | 18 | 1156 [1287] | 82.97% | 46 [64] | 323 [386] | 26.08% |
| 34 | 48 | 1158 [1685] | 85.23% | 56 [78] | 724 [1173] | 55.12% |
| 43 | 62 | 1158 [1914] | 86.22% | 62 [88] | 848 [1549] | 64.31% |
| 86 | 135 [136] | 1158 [3560] | 91.38% | 68 [160] | 1102 [3630] | 82.68% |
| 111 | 192 | 1158 [4370] | 95.41% | 82 [228] | 1114 [4802] | 84.53% |

**Table 9. LOM vs STORM coverage with lower $rmin_c$.** LOM compared to STORM coverage (%) and how the coverage distributes between low- and high-priority grid cells (in number of cells looked at of each kind). Numbers in brackets give total number of looks if higher than the number of grid cells looked at.

We next adjust the threshold resolution $rmin_c$ (uniformly implemented for all grid cells) as specified in Table 10. As the threshold increases, the coverage naturally drops as fewer grid cells can be looked at with high resolution. In these instances, the STORM heuristic performs on par or only moderately worse than LOM. (We note that 262 and 1,171 grid cells in row one, columns five and six seem to exceed the available grid cells (257 and 1,158, respectively). This is attributed to our center-point method of calculating when a grid cell overlaps with a swath. We suspect that STORM determines this differently.)

| $rmin_c$ | LOM | | | STORM heuristic | | |
|---|---|---|---|---|---|---|
| | High priority | Low priority | Coverage | High priority | Low priority | Coverage |
| 1 | 257 [753] | 1153 [2130] | 99.64% | 262 [697] | 1171 [2136] | 100% |
| 2 | 257 | 680 [725] | 66.22% | 227 [364] | 583 [616] | 57.24% |
| 3 | 11 | 495 [508] | 35.76% | 157 [241] | 278 [279] | 30.74% |
| 4 | 0 | 135 | 9.54% | 55 [76] | 64 | 8.41% |
| 5 | 0 | 136 | 9.61% | 21 [30] | 80 | 7.14% |

**Table 10. LOM vs STORM coverage with varying $rmin_c$.** LOM compared to STORM coverage (%) and how the coverage distributes between low- and high-priority grid cells (in number of cells looked at of each kind). Numbers in brackets give total number of looks if higher than the number of grid cells looked at. We require all grid cells to use the specified $rmin_c$ except in the last row ($rmin_c = 5$). In that case, we set $rmin_c = 5$ for high-priority and $rmin_c = 4$ for low-priority grid cells to allow for use of sensors that do not have resolution 5.

We examine further the resolution levels recommend by LOM. Table 11 provide results when threshold resolution $rmin_c = 3$ for high-priority grid cells and $rmin_c = 1$ for low-priority ones.

LOM achieves 99% coverage, while the STORM heuristic achieves 77.1%. In this instance, LOM covers high-priority grid cells almost as well as low-priority ones. With the focus on coverage, LOM has little incentive to recommend a resolution level above the threshold.

| Resolution | LOM | | | STORM heuristic | | |
|------------|-----|-----|-----|-----|-----|-----|
| $r$ | High priority | Low priority | Total | High priority | Low priority | Total |
| 1 | 0 | 1157 [1624] | 1157 [1624] | 0 | 934 [1460] | 934 [1460] |
| 2 | 0 | 1 [18] | 1 [18] | 0 | 0 | 0 |
| 3 | 239 | 0 [1] | 239 [240] | 142 [218] | 0 | 142 [218] |
| 4 | 4 | 0 | 4 | 15 [20] | 0 | 15 [20] |
| 5 | 0 | 0 | 0 | 0 | 0 | 0 |
| Total | 243 | 1158 [1633] | 1401 [1876] | 157 [238] | 934 [1460] | 1091 [1698] |
| Coverage | 94.6% | 100% | 99.0% | 61.1% | 80.6% | 77.1% |

**Table 11. LOM vs STORM coverage; resolution detail**. LOM compared to STORM coverage (%) and how the coverage distributes between low- and high-priority grid cells (in number of cells looked at of each kind). Numbers in brackets give total number of looks if higher than the number of grid cells looked at. We use $rmin_c = 3$ for high-priority grid cells and $rmin_c = 1$ for low-priority ones.

The flexibility of LOM is highlighted by changing *maxlow* from 0 to 1 and thus allow for one look with low resolution (below the threshold $rmin_c$). This may emerge as a useful setting for planners that seek high coverage, but also high resolution when possible. To explore this possibility, we set $rmin_c = 2$ for all grid cells and obtain the results in Table 12. LOM achieves 100% coverage using mostly a coarse resolution, with some additional higher resolution looks. The STORM heuristic does not have flexibility of this kind and needs to stick to preselected resolution levels.

| Resolution $r$ | LOM | | | STORM heuristic | | |
|---|---|---|---|---|---|---|
| | High priority | Low priority | Total | High priority | Low priority | Total |
| 1 | 31 [257] | 1067 [1158] | 1157 [1624] | 0 | 0 | 0 |
| 2 | 226 [324] | 88 [308] | 1 [18] | 211 [344] | 571 [600] | 782 [944] |
| 3 | 0 | 3 [11] | 239 [240] | 0 | 0 | 0 |
| 4 | 0 | 0 [5] | 4 | 16 [20] | 12 [16] | 28 [36] |
| 5 | 0 | 0 [1] | 0 | 0 | 0 | 0 |
| Total | 257 [581] | 1158 [1633] | 1158 [1483] | 227 [364] | 583 [616] | 810 [980] |
| Coverage | 100% | 100% | 100% | 88.32% | 50.35% | 57.24% |

**Table 12. LOM vs STORM coverage with $maxlow = 1$.** LOM compared to STORM coverage with $rmin_c = 2$ for all grid cells and $maxlow = 1$. We report how the coverage distributes between low- and high-priority grid cells (in number of cells looked at of each kind). Numbers in brackets give total number of looks if higher than the number of grid cells looked at.

## CONCLUSIONS

In this paper, we develop LOM, which is a mixed-integer linear optimization model that seeks to distribute a limited number of looks by satellite-borne sensors to grid cells while accounting for various constraints. We develop the model in the context of the campaign simulation tool STORM, which also generates realistic test instances.

Through tuning of options in CPLEX, we overcome solver failure (no feasible solution after 1,400 minutes) and achieve quality allocation plans in less than 5 minutes. In large-scale instances with 22 million variables and 11 million constraints, we achieve an optimality gap of 9% in less than 40 minutes. Across 28 operational case studies, LOM provides an average of 54.6% and a median of 22.8% more coverage of grid cells compared to a heuristic algorithm in STORM. LOM presents planners with a tool for balancing concerns about coverage against the need for timely high-resolution images.

Instances of LOM with long planning horizons exhibit variation in runtimes, with some exhausting the memory of a 128-GB desktop computer before reaching a desired 5% optimality gap. An immediate remedy is to adopt problem cascade techniques (see, e.g., Brown et al., 1987) or simply to implement a rolling horizon approach. LOM can be extended to account for non-satellite-borne sensors, but this requires significant changes to the model because airborne sensors must obey travel-time constraints as they move between grid cells. There is also a need for considering the uncertainty in cloud cover and other factors, which leads to stochastic optimization problems (see, e.g., Chapter 3 of Royset and Wets, 2021) and new sets of modeling challenges.

## ACKNOWLEDGEMENTS

This work is supported in part by the Office of Naval Research (Mathematical and Resource Optimization Program) under MIPRN0001422WX01445.